\documentclass[structabstract]{aa}  
\usepackage{graphicx}
\usepackage{txfonts}
\usepackage{natbib}

\usepackage{longtable}
\usepackage{comment}

\usepackage{multirow}

\begin{document}
   \title{The evolving spectrum of the planetary nebula Hen\,2-260}

  \authorrunning{M. Hajduk et al.}

   \author{M. Hajduk\inst{1}, P. A. M. van Hoof\inst{2}, K. Gesicki\inst{3}, A. A. Zijlstra\inst{4}, S. K. G\'{o}rny\inst{1}, and M. G{\l}adkowski\inst{1}}

   \institute{Nicolaus Copernicus Astronomical Center, 
ul. Rabia\'{n}ska 8, 87-100 Toru\'{n}, Poland \and
    Royal Observatory of Belgium, Ringlaan 3, 1180 Brussels, Belgium \and
    Centre for Astronomy, Faculty of Physics, Astronomy and Informatics,
    Nicolaus Copernicus University, Grudziadzka 5, PL-87-100 Torun, Poland \and
    Jodrell Bank Centre for Astrophysics, Alan Turing Building,
Manchester M13 9PL, UK}

   \date{Received ; accepted }

  \abstract{}
{We analysed the planetary nebula Hen\,2-260 using optical spectroscopy and photometry. We compared our observations with the data from literature to search for evolutionary changes. We also searched for photomertic variability of the central star.}
{The object Hen\,2-260 was observed with the SAAO 1.0m telescope (photometry) and the SALT telescope (low resolution spectroscopy). We also used archival high resolution Very Large Telescope spectra and Hubble Space Telescope imaging. The nebular line fluxes were modelled with the Cloudy photoionization code to derive the stellar and nebular parameters.}
{The planetary nebula shows a complex structure and possibly a bipolar outflow. The nebula is relatively dense and young. The central star is just starting $\rm O^+$ ionization ($\rm T_{eff} \approx 30,000 \, K$). Comparison of our observations with literature data indicates a 50\% increase of the [O\,{\sc iii}] 5007\,\AA\ line flux between 2001 and 2012. We interpret it as the result of the progression of the ionization of $\rm O^{+}$. The central star evolves to higher temperatures at a rate of $\rm 45 \pm 7\,K\, \rm yr^{-1}$. The heating rate is consistent with a final mass of $\rm 0.626 ^{+0.003} _{-0.005} \, M_{\odot}$ or $\rm 0.645 ^{+0.008} _{-0.008} \, M_{\odot}$ for two different sets of post-AGB evolutionary tracks from literature. The photometric monitoring of Hen\,2-260 revealed variations on a timescale of hours or days. There is no direct indication for central star binarity in the spectrum nor for a strong stellar wind. The variability may be caused by pulsations of the star.}
{The temperature evolution of the central star can be traced using spectroscopic observations of the surrounding planetary nebula spanning a timescale of roughly a decade. This allows us to precisely determine the stellar mass, since the pace of the temperature evolution depends critically on the core mass. The method is independent of the absolute age of the nebula. The kinematical age of the nebula is consistent with the age obtained from the evolutionary track. The final mass of the central star is close to the mass distribution peak for central stars of planetary nebulae found in other studies. The object belongs to a group of young central stars of planetary nebulae showing photometric variability.}

\maketitle
%

\section{Introduction}

Planetary nebulae (PNe) are the final stage in evolution of stars with initial
masses of 1-8 $\rm\,M_\odot$. Low-mass stars lose most of their material via a
stellar wind during the asymptotic giant branch (AGB) phase. Once the star has
shed its envelope, only a hot core surrounded by a tiny ($\rm
10^{-4}\,M_{\odot}$) hydrogen envelope remains. The star subsequently moves
towards the white dwarf regime in the Herzsprung-Russell diagram, where the
thermonuclear reactions finally stop.

The pace of evolution critically depends on the final mass of the central star
of a PN. The star reaches its maximum temperature on a timescale of several
decades for the most massive objects up to several thousands years for low mass
central stars \citep{1995A&A...299..755B}. On its path towards high effective
temperatures, the central star passes through the instability strip. Pulsational
instabilities in post-AGB stars have been studied by \citet{1993AcA....43..431Z}
and \citet{1993MNRAS.265..340G}. Several examples of variable young central
stars of PNe have been found by \citet{2003IAUS..209..237H}. Variability was
attributed to either stellar wind instabilities or pulsations.

Only in a few cases, a change of the ionizing flux of a central star was
observed on a timescale of years or decades. \citet{1992PASP..104..339F}
reported a variability of emission lines in the PNe IC\,4997 and NGC\,6572. More
recently, \citet{2013AstL...39..201A} discovered the variability of the PN
Hen\,3-1357 and its central star. In the case of NGC\,7027, observations at
radio frequencies for 25 years allowed for the determination of the central star
mass \citep{2008ApJ...681.1296Z}.

\section{PN Hen\,2-260}

The planetary nebula Hen\,2-260 (PN\,G 008.2+06.8) is located 10 degrees away
from the Galactic Center. \citet{2007A&A...467.1253H} modelled the central star
of Hen\,2-260 with a H-rich model atmosphere with $\rm T_{eff} = 28,000 \pm
1500$\,K. No stellar emission lines were observed, which indicated a low mass
loss rate. \citet{2007A&A...467.1253H} derived a spectroscopic distance to the
nebula of $11.2^{+2.0}_{-1.6}$ kpc. Most of the statistical determinations of
the distance in the literature for Hen\,2-260 are consistent with this result
(e.g., 12.17 kpc derived by \citealp{1995ApJS...98..659Z}; 11.44 kpc by
\citealp{1994A&AS..108..485V}; $17.0\pm 3.4$ kpc by
\citealp{2010ApJ...714.1096S}).

The dust shows the presence of amorphous and crystalline silicates
\citep{2009A&A...495L...5P,2010A&A...516A..39G}. \citet{2011MNRAS.414.1667G}
discovered very faint  PAH emission at $8.6 \, \mu m$.

An excitation class of 0.5 was given to this object by
\citet{2004A&A...414..211E}. They determined the (reddened) [O\,{\sc iii}]
5007\,\AA\ line flux to be only 5.1 (Table \ref{ratios}).
\citet{1991A&AS...89..237A} did not observe the [O\,{\sc iii}] 5007\,\AA\
nebular line in the spectrum taken almost two decades before.

\section{Observations}

The photometric observations were done at the SAAO observatory using the 1.0m
telescope in five periods between 2010 August 25 -- September 1, 2011 March
24--29, 2011 May 4--10, 2011 August 24--30, and 2012 May 9--22 in the I filter.
We used the SITe CCD camera, which measures 1024 $\times$ 1024 pixels with the
field of view of about 5 arcmin squared.

The Hubble Space Telescope (HST) snapshot survey of bulge PNe was held in 2002
and 2003 (proposal 9356, PI Albert Zijlstra). The images were obtained using the
Wide Field Planetary Camera in the F656N (two 100-s exposures), F502N (two 230-s
exposures), and F547M (one 60-s exposure) filters on 2003 March 13.

High-resolution spectra were obtained with the Very Large Telescope (VLT) on
2005 April 21 with the Ultraviolet and Visual Echelle Spectrograph (UVES) in
three wavelength ranges: 3280--4480, 4600--5600, and 5660--6660\,\AA\ with the
spectral resolution of about $\lambda / \Delta \lambda = 60\,000$. Three 600-s
exposures were made. The slit was 11 arcsec long and 0.5 arcsec wide. The
spectrum was flux calibrated using a standard star observation. The line flux
uncertainty derived from the comparison of the fluxes of the same lines measured
in individual spectra is about 2\% for the strong lines and 6\% for the lines
$\leq 1\%$ of the $\rm H\,\beta$ flux.


The SALT low-resolution spectra were obtained on 2012 May 29 with the 900\,l/mm 
grating. The 1 arcsec wide and 8 arcmin long slit was used. The spectral
coverage was 4350--7400\,\AA\ with a spectral resolution of about 6000 at the
central wavelength. Two spectra were taken with exposure times of 120 and 900
sec. Most of the fluxes used in this work were derived from the 900 sec
spectrum, except for the $\rm H\alpha$, $\rm H\beta$, [N\,{\sc ii}]\,6548\,\AA,
[N\,{\sc ii}]\,6584\,\AA, [O\,{\sc ii}]\,7320\,\AA\ and [O\,{\sc
ii}]\,7330\,\AA\ lines. These lines were saturated in the 900 sec spectrum and
their fluxes were measured in the 120 sec spectrum.

The spectra were flux calibrated using the standard star observation taken
during the same night. Line fluxes were measured from both spectra and compared.
The line flux uncertainty derived from the comparison of 120 and 900 sec spectra
is about 5\% for the strong lines, 10\% for the lines $\leq 1\%$ of the $\rm
H\,\beta$ flux, and 25\% for $\leq 0.2\%$ of the $\rm H\,\beta$ flux.

The Radcliffe 1.9\,m telescope low-resolution spectra were obtained using the
grating spectrograph with a 1.5 arcsec slit between 2012 May 5 and 9. The total
exposure time was about 3 hours. The wavelength coverage is from 3625 to
7600\,\AA\ and the spectral resolution of about 2500 at the central wavelength.
The spectra were flux calibrated. A different set of flux standards was measured
during each night. The uncertainty of the fluxes is about 5\%, measured from the
spectra collected in different nights.

Some of the nebular fluxes may be affected by the underlying stellar
absorptions. This applies to faint He\,{\sc i} absorption lines in the
low-resolution spectra. The He\,{\sc i} fluxes may be systematically higher in
the VLT spectrum than in the low resolution spectra.

The data were reduced using the {\sc iraf} package. We used the {\sc isis}
difference image analysis package to obtain magnitudes. This allows us to avoid
the instrumental effects caused by the presence of a resolved nebula. Otherwise,
the nebular contribution may degrade the quality of the lightcurve.

To search for periodic variability, we used the {\sc period04} program
\citep{2005CoAst.146...53L} and the analysis of variance (ANOVA) method
\citep{1996ApJ...460L.107S}.

\section{Results}

\subsection{HST imaging}\label{hst}

\begin{figure}[ht]
\begin{center}
\includegraphics[width=4.5cm]{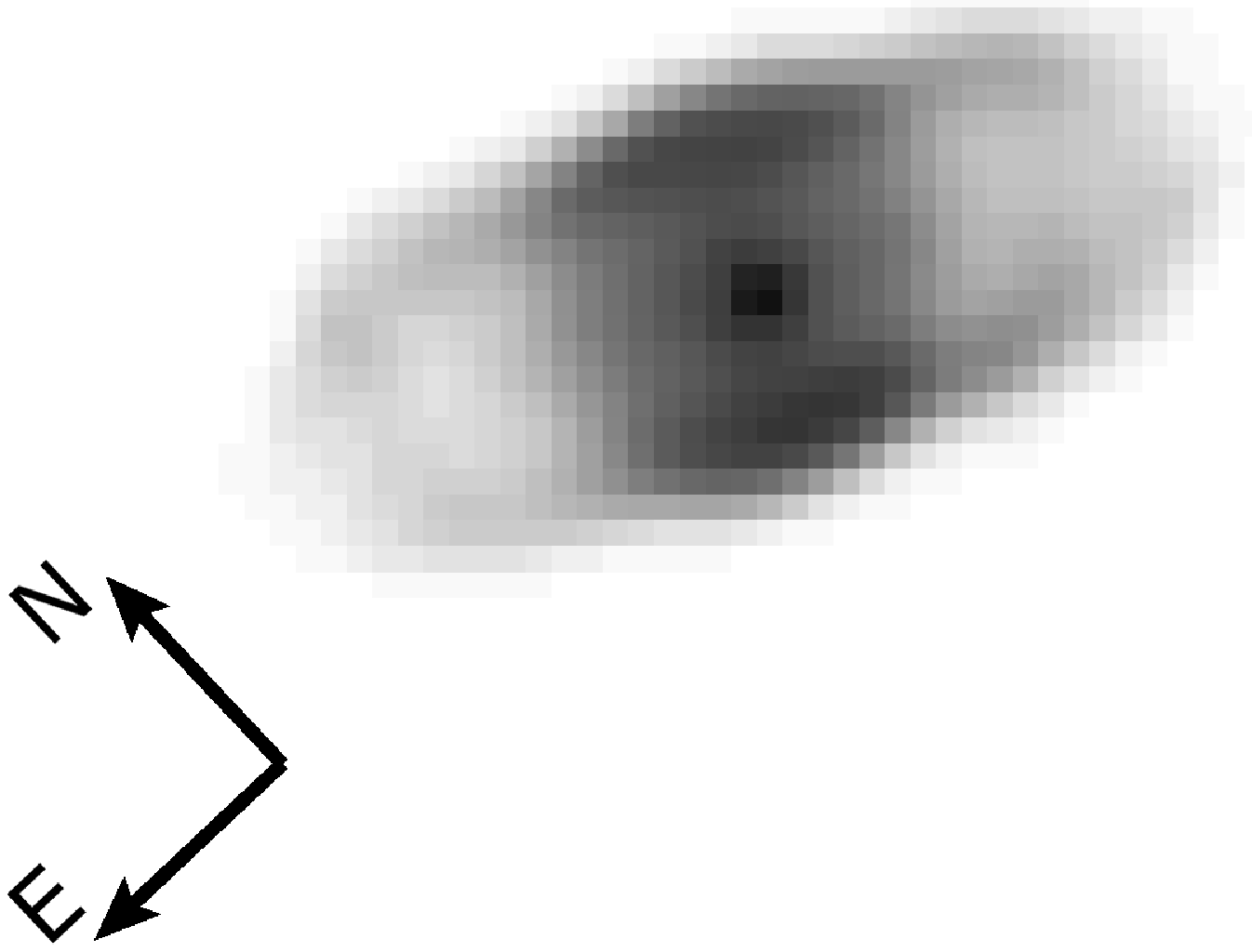}\includegraphics[width=4.5cm]{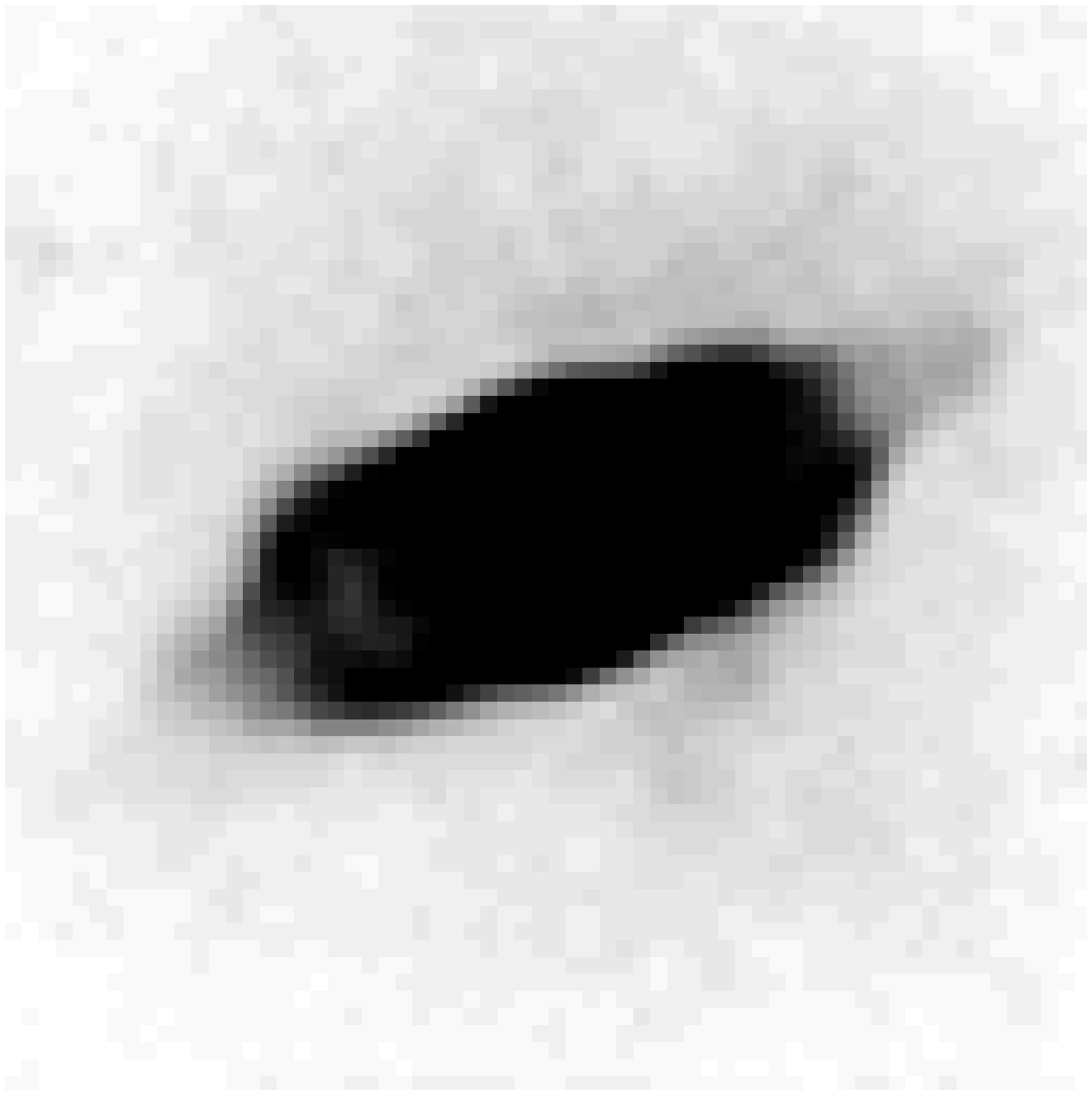}

\noindent\includegraphics[width=4.52cm]{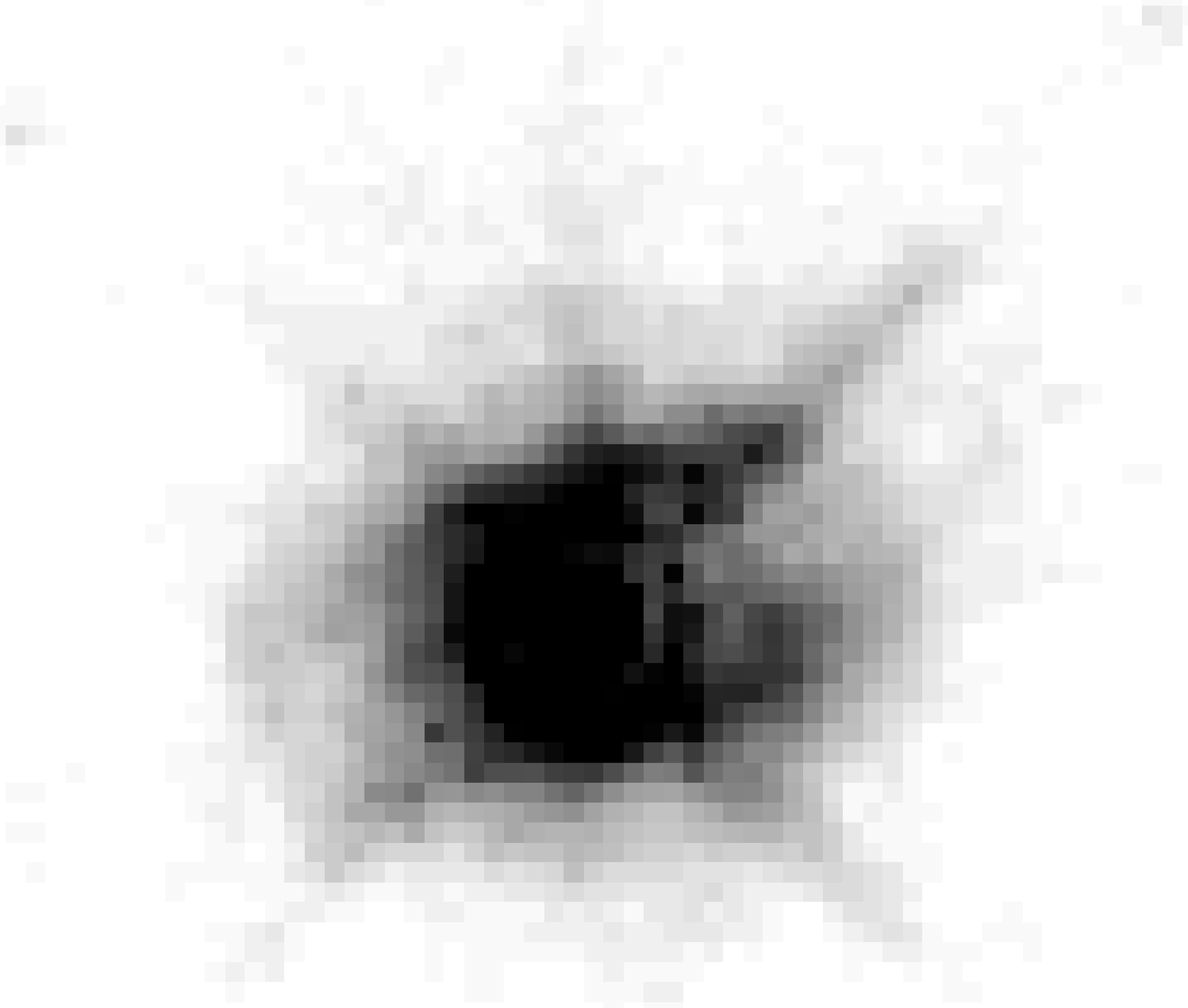}\includegraphics[width=4.6cm]{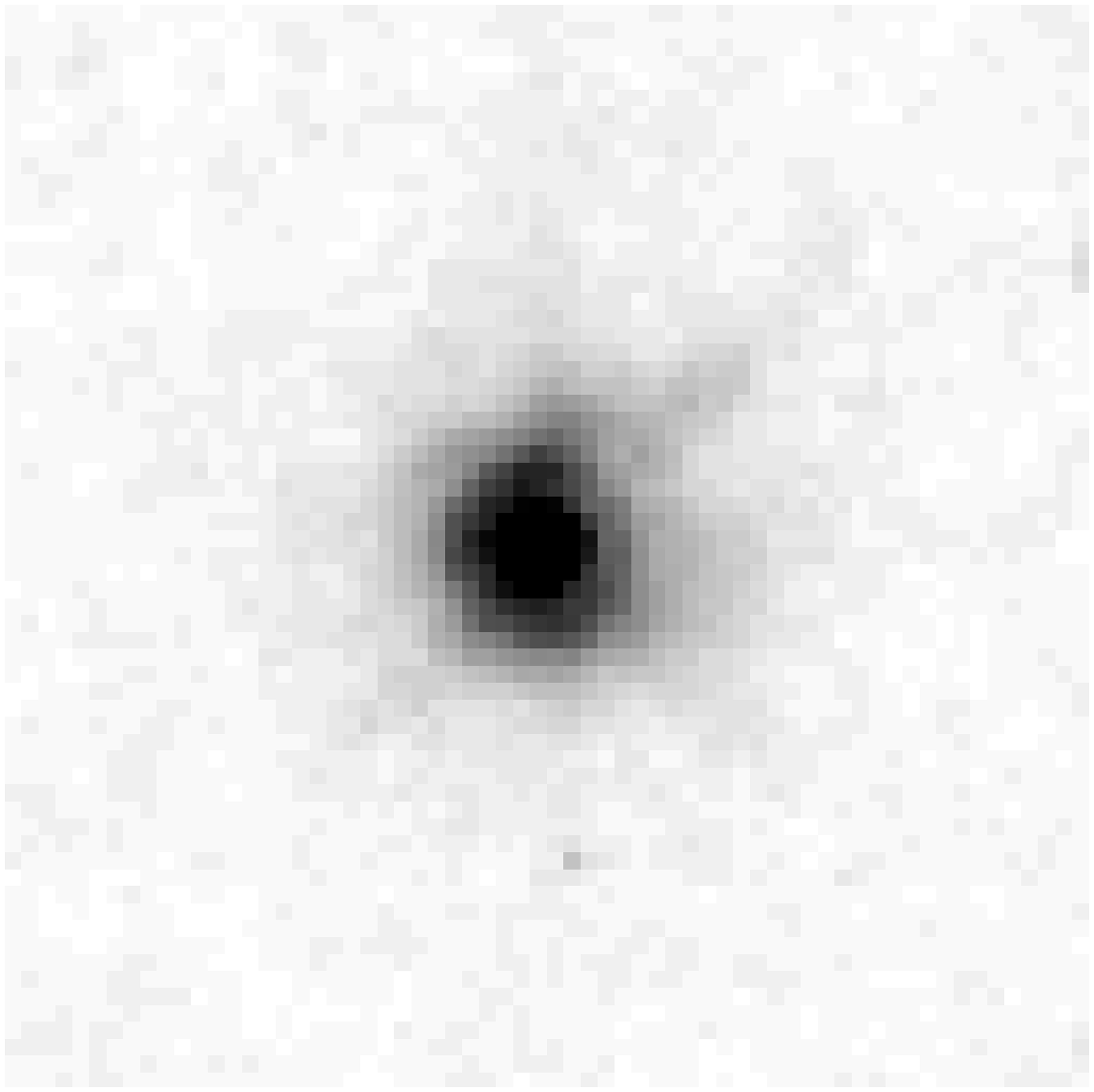}
\end{center}
\caption{Upper row: the HST F656N deconvolved image of the PN Hen\,2-260. Different cut levels were applied to enhance the bipolar structure (left) and bipolar outflow (right). Lower row: the HST F547M (left) and F502N (right) deconvolved images. All the images are 3 arcsec wide. Logarithmic intensity scale was applied to all the images.}
\label{hst}
\end{figure}

The HST image of Hen\,2-260 reveals a complex morphology of the object. The
brightest, central part of the nebula appears to be roughly spherical. An outer,
elongated limb-brightened structure appears to be open near the polar axes
(Figure \ref{hst}). A polar outflow appears to be present in the F656N ($\rm H
\alpha$) filter image. The maximum extent of the nebula is $1.8 \times 0.85$
arcsec in the F656N filter image.

The central star is clearly visible in the F547M (continuum) and F502N ([O\,{\sc
iii}] 5007\,\AA) filters. The V-band magnitude of the central star, measured
using the F547M filter image, is $14.13 \pm 0.03$. We observe a faint, extended
nebular component in the F547M filter image (and probably in F502N, too) after
the deconvolution was performed with the point spread function (PSF) that was
generated with the TinyTim tool \citep{2011SPIE.8127E..16K}. However, a residual
PSF pattern (diffration spikes and the Airy rings) is clearly present in the
images and hampers the analysis of the F547M and F502N filter images. The
brightest parts of the nebula in the $\rm H\alpha$ image are located at a
projected distance of about 0.2 arcsec away from the central star, which is
similar to the radius of to the 1st Airy ring.

We obtained ${\rm log \,}F({\rm H\alpha}) {\rm = -11.19 \pm 0.03 \, erg \,
cm^{-2} \, s^{-1}}$ from the HST image, using the calibration formulae given by
\citet{1997hstc.work..338D}. The $\rm H \alpha$ line equivalent width is about
$-670\,\AA$, while the F656N filter width is 21.5\,\AA. Thus, the continuum
contribution to the F656N inband flux is only about 3\%.

We obtained an [O\,{\sc iii}] 5007\,\AA\ to $\rm H \alpha$ line flux ratio of
0.07 from the HST images. The [O\,{\sc iii}] 5007\,\AA\ to $\rm H \alpha$ line
flux ratio obtained from the SALT spectrum is about 0.016. The discrepancy is
due to the continuum contribution to the F502N filter flux. The equivalent width
of the [O\,{\sc iii}] 5007\,\AA\ line is $-8\,\AA$ (measured from the SALT
spectrum), while the F502N filter width is 26.9\,\AA. The contribution of the
continuum to the flux in the F502N filter is expected to be 3.4 times higher
than the [O\,{\sc iii}] 5007\,\AA\ line flux. The contribution of the [Fe\,{\sc
iii}] 5011\,\AA\ and He\,{\sc i} 5016\,\AA\ flux to the inband flux in the F502N
filter is about 10\% of the [O\,{\sc iii}] 5007\,\AA\ line flux.

\subsection{Optical spectra}

\begin{table*}
\caption{Comparison of fluxes of the strongest nebular lines.}
\begin{center}
\begin{tabular}{ccccccc}
\hline
$\lambda \, [\AA]$ &ion&ESO 1.52m$^{\mathrm{a}}$&ESO 1.52m$^{\mathrm{b}}$&VLT &SAAO 1.9\,m &SALT\\
	&	&1984-04-30&2001-07-01&2005-04-20&2012-05-07&2012-05-29\\
	&	&JD\,2445820   &JD\,2452091   &JD\,2453480   &JD\,2456054   &JD\,2456076   \\
\hline
4861	&H\,{\sc i}	&$100.0\pm5.0$	&$100.0\pm2.7$	&$100.0\pm1.0$	&$100.0\pm10.0$	&$100.0\pm5.0$	\\
4959	&[O\,{\sc iii}]	&-		&$1.6\pm0.2$	&$1.8\pm0.1$	&$2.6\pm0.3$	&$2.4\pm0.3$	\\
5007	&[O\,{\sc iii}]	&-		&$5.1\pm0.4$	&$5.5\pm0.1$	&$7.8\pm0.8$	&$7.9\pm0.4$	\\
5754	&[N\,{\sc ii}]	&$3.0\pm1.0$	&$4.1\pm0.4$	&$4.1\pm0.1$	&$4.1\pm0.5$	&$4.2\pm0.3$	\\
5875	&He\,{\sc i}	&$4.0\pm1.0$	&$2.1\pm0.2$	&$2.8\pm0.1$	&$2.9\pm0.3$	&$2.9\pm0.2$	\\
6300	&[O\,{\sc i}]	&$5.0\pm1.0$	&$3.5\pm0.3$	&$3.4\pm0.1$	&$3.7\pm0.4$	&$3.4\pm0.2$	\\
6312	&[S\,{\sc iii}]	&-		&$1.3\pm0.2$	&$1.4\pm0.1$	&$2.0\pm0.2$	&$1.5\pm0.1$	\\
6364	&[O\,{\sc i}]	&$2.2\pm0.7$	&$1.1\pm0.2$	&$1.1\pm0.1$	&$1.1\pm0.2$	&$1.1\pm0.1$	\\
6548	&[N\,{\sc ii}]	&$77.0\pm4.0$	&$59.6\pm1.9$	&$57.8\pm0.6$	&$61.8\pm6.2$	&$62.9\pm3.2$	\\
6563	&H\,{\sc i}	&$581.0\pm30.0$	&$498.9\pm7.7$	&$509.0\pm5.0$	&$493.0\pm50.0$	&$493.0\pm25.0$	\\
6584	&[N\,{\sc ii}]	&$213.0\pm11.0$	&$183.8\pm4.0$	&$190.0\pm2.0$	&$179.0\pm18.0$	&$186.0\pm10.0$	\\
6678	&He\,{\sc i}	&-		&$0.5\pm0.1$	&		&$1.4\pm0.2$	&$0.8\pm0.1$	\\
6716	&[S\,{\sc ii}]	&$4.0\pm1.0$	&$4.2\pm0.4$	&		&$4.2\pm0.5$	&$3.6\pm0.4$	\\
6731	&[S\,{\sc ii}]	&$10.0\pm1.0$	&$9.0\pm0.6$	&		&$8.0\pm0.8$	&$8.1\pm0.5$	\\
7002	&[O\,{\sc i}]	&-		&$0.9\pm0.2^{\mathrm{c}}$& 	&$0.9\pm0.1$	&$0.7\pm0.1$	\\
7065	&He\,{\sc i}	&-		&$1.1\pm0.2$	&		&$1.9\pm0.2$	&$1.9\pm0.1$	\\
7135	&[Ar\,{\sc iii}]&-		&$1.6\pm0.2$	&		&$2.1\pm0.3$	&$2.3\pm0.2$	\\
7320	&[O\,{\sc ii}]	&\multirow{2}{*}{ {\Large \}} $80.0\pm4.0$} &$35.8\pm1.5$&&$44.8\pm2.3$	&$49.8\pm2.5$	\\
7330	&[O\,{\sc ii}]	& {}	&$41.2\pm1.8$	&&$39.5\pm2.0$	&$40.4\pm2.1$	\\
\hline	
\label{ratios}
\end{tabular}
\end{center}
$^{\mathrm{a}}$\citet{1991A&AS...89..237A} $^{\mathrm{b}}$\citet{2004A&A...414..211E} $^{\mathrm{c}}$misidentified as [Ar\,{\sc v}] $\lambda 7005.6$ by \citet{2004A&A...414..211E}
\end{table*}

We obtained the extinction of $\rm c(H\beta) = 0.69 \pm 0.09$ by using the $\rm
H\alpha$ to $\rm H\beta$ line flux ratio and by assuming the standard value of
the total to selective absorption ratio $\rm R_V = \frac{A_V}{E(B-V)}$ of 3.1.
The optical spectra were dereddened using the extinction curve given by
\citet{1999PASP..111...63F}.

The ratio of the radio continuum flux at 5\,GHz and the observed $\rm H \beta$
flux (both quantities are given in Table \ref{parameters}) can also be used to
obtain the extinction (e.g., \citet{1984ASSL..107.....P}). The predicted
(dereddened) ${\rm log}\,F{\rm (H\beta)}$ flux is $\rm
-11.37\,erg\,cm^{-2}\,s^{-1}$, which gives $\rm c(H\beta)$ of $0.76 \pm 0.11$. 

\citet{2013A&A...550A..35P} obtained the extinction $\rm c(H\beta)$ of 1.06 by
using the H(6-5) $7.46 \, \mu m$ and H(7-6) $12.37 \, \mu m$ line fluxes. They
suggested that the discrepancy between the extinction obtained by using radio
continuum flux and the infrared hydrogen line fluxes could result from the
non-negligible optical depth at radio frequencies. However, Hen\,2-260 has a
brightness temperature of only 920\,K at 5\,GHz \citep{1990A&AS...84..229A}.
Using the emission measure computed in the Cloudy model, we determined the
optical thickness of $\tau_{1.4GHz}=0.45$ and $\rm \tau_{5GHz} = 0.03$. The
observed fluxes of the radio spectrum are consistent with those results. The
agreement of the extinction determinations using the optically thin 5\,GHz with
the extinction from the Balmer decrement validates our choice.

The nebular line fluxes are shown in Table \ref{ratiossalt} (SALT and 1.9\,m
Radcliffe telescope spectra) and Table~4\footnote{Table 4 is only available in
electronic form at the CDS via anonymous ftp to cdsarc.u-strasbg.fr
(130.79.128.5) or via http://cdsweb.u-strasbg.fr/cgi-bin/qcat?J/A+A/} (VLT
spectrum). We compared the observed and theoretical ratios of the pairs of lines
coming from the same upper levels to assess the reliability of the error
determinations of the nebular fluxes. The transition probabilities were taken
from the following: \citet{1996A&AS..119..509N} for [Fe\,{\sc iii}] lines,
\citet{1996A&AS..120..361Q} for [Fe\,{\sc ii}] lines and 
\citet{2000MNRAS.312..813S} for [O\,{\sc iii}] 5007/4959\,\AA, [N\,{\sc ii}]
6584/6548\,\AA\ and [O\,{\sc i}] 6300/6364\,\AA\ line ratios. The Si\,{\sc ii}
transition probabilities were taken from the Atomic Line
List\footnote{http://www.pa.uky.edu/$\sim$peter/newpage/}.

The average $\chi ^2$ was 0.38 for the six ratios determined from the SALT
spectrum and 1.32 for the VLT spectrum for the nineteen ratios. We assumed a
20\% uncertainty for the transition probabilities for the [Fe\,{\sc iii}] lines,
15\% uncertainty for the [Fe\,{\sc ii}] lines, 6\% to 15\% uncertainty for the
Si\,{\sc ii} lines, and negligible uncertainty for the remaining lines. The main
sources of the high value of $\chi ^2$ derived for the VLT spectrum were the
[N\,{\sc ii}] 6584/6548\,\AA\ line flux ratio and the Si\,{\sc ii}
5958/5979\,\AA\ line ratio. The 6584\,\AA\ line flux can be affected by
uncertain calibration close to the edge of the wavelength range. The Si\,{\sc
ii} 5958\,\AA\ line flux can be affected by the nearby O\,{\sc i} emission. We
obtained an averaged $\chi ^2$ of 0.80 for seventeen remaining flux ratios. Good
agreement between the theoretical and observed flux ratios confirms that we
derived reliable uncertainties for the fluxes. 

We compared the fluxes derived with the SALT telescope with the fluxes published
by \citet{2004A&A...414..211E}. Some of the fluxes agree very well (Table
\ref{ratios}). However, [O\,{\sc iii}], He\,{\sc i}, [Ar\,{\sc iii}] and
[S\,{\sc iii}] fluxes seem to have increased between 2001 and 2012. In
particular, we obtained a [O\,{\sc iii}] 5007\,\AA\ line flux of $7.9 \pm 0.4$
(relative to F($\rm H\beta) = 100$) using the SALT spectrum, which is markedly
higher than previous observers. \citet{2004A&A...414..211E} obtained $5.1 \pm
0.4$, while \citet{1991A&AS...89..237A} do not report a detection of this line.
The flux increase may indicate the increasing stellar temperature.

One can question if the observed increase in the [O\,{\sc iii}] 5007\,\AA\ flux
is not of instrumental origin. The spectra in different epochs were obtained
with different instruments and configurations. In particular,
\citet{2004A&A...414..211E} used a 2 arcsec wide slit; a 1 arcsec slit was used
in the SALT spectrum; a 1.5 arcsec slit for the 1.9\,m Radcliffe telescope
spectrum; and the VLT spectrum was obtained with a 0.5 arcsec slit.

The spectra obtained more recently could be more sensitive to the inner regions
of the nebula compared to the spectra obtained before, due to light loss on the
slit. This could affect the observed [O\,{\sc iii}] 5007\,\AA\ to $\rm H\beta$
line flux ratio.

The spectrum published by \citet{2004A&A...414..211E} was obtained with a 2
arcsec wide slit, which did not resolve the nebula. The SALT spectrum was
obtained with a 1 arcsec wide slit, which could resolve the nebula, but only
under very good seeing conditions. However, the seeing during the SALT
observation was about 1.3 arcsec, a value that is comparable to the extent of
the nebula. The seeing during the 1.9\,m Radcliffe telescope observations, which
were obtained in the same year, varied between 1.3 and 1.9 arcsec. The SALT and
1.9\,m Radcliffe telescope spectra show a consistent [O\,{\sc iii}] 5007\,\AA\
flux, a value that is higher than in the spectrum by \citet{2004A&A...414..211E}
and also higher than in the VLT spectrum, which was obtained with the most
narrow slit of all the observations used in this work. This confirms that the
flux increase is real.

\begin{figure}
\centering
\includegraphics[width=8cm]{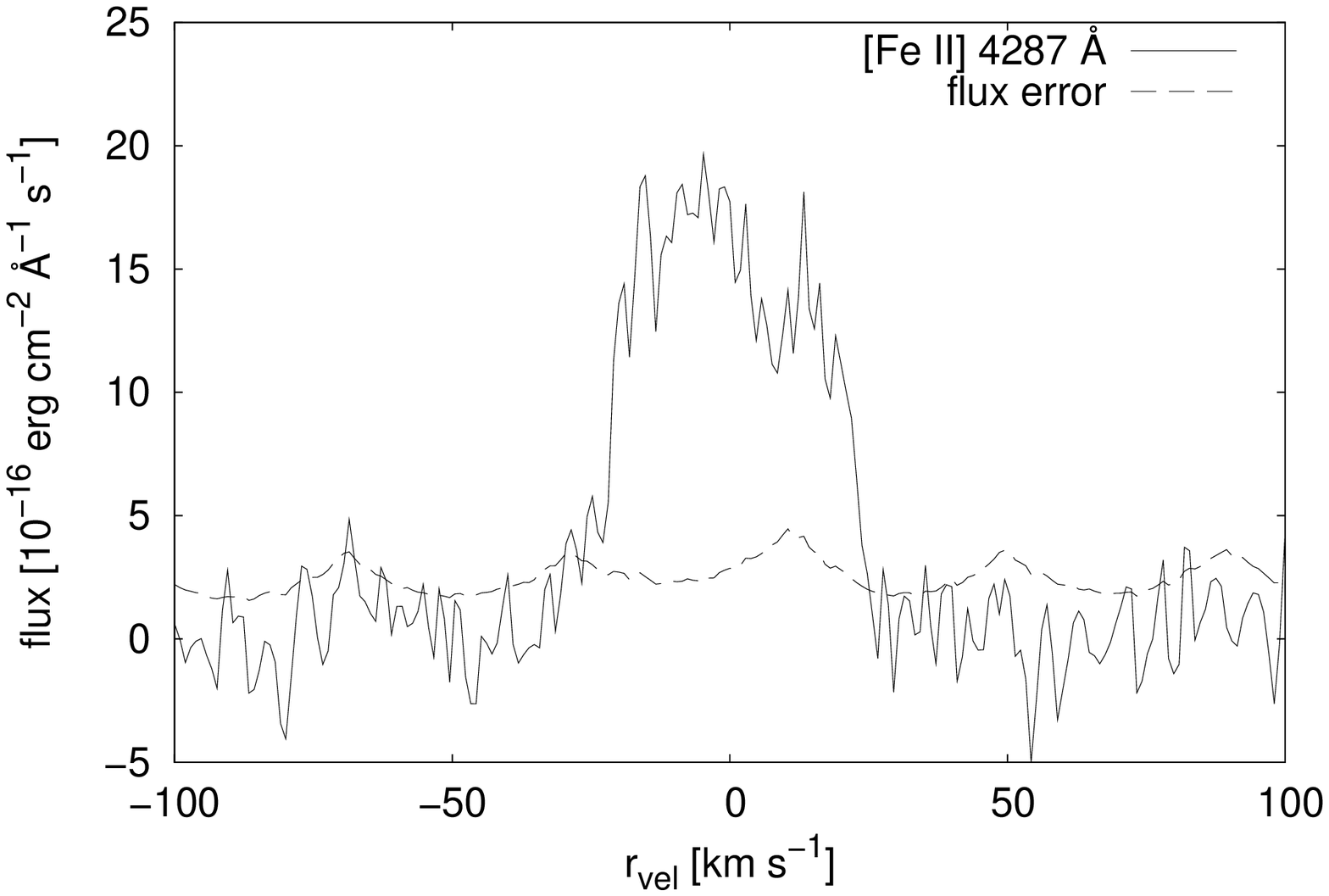}
\includegraphics[width=8cm]{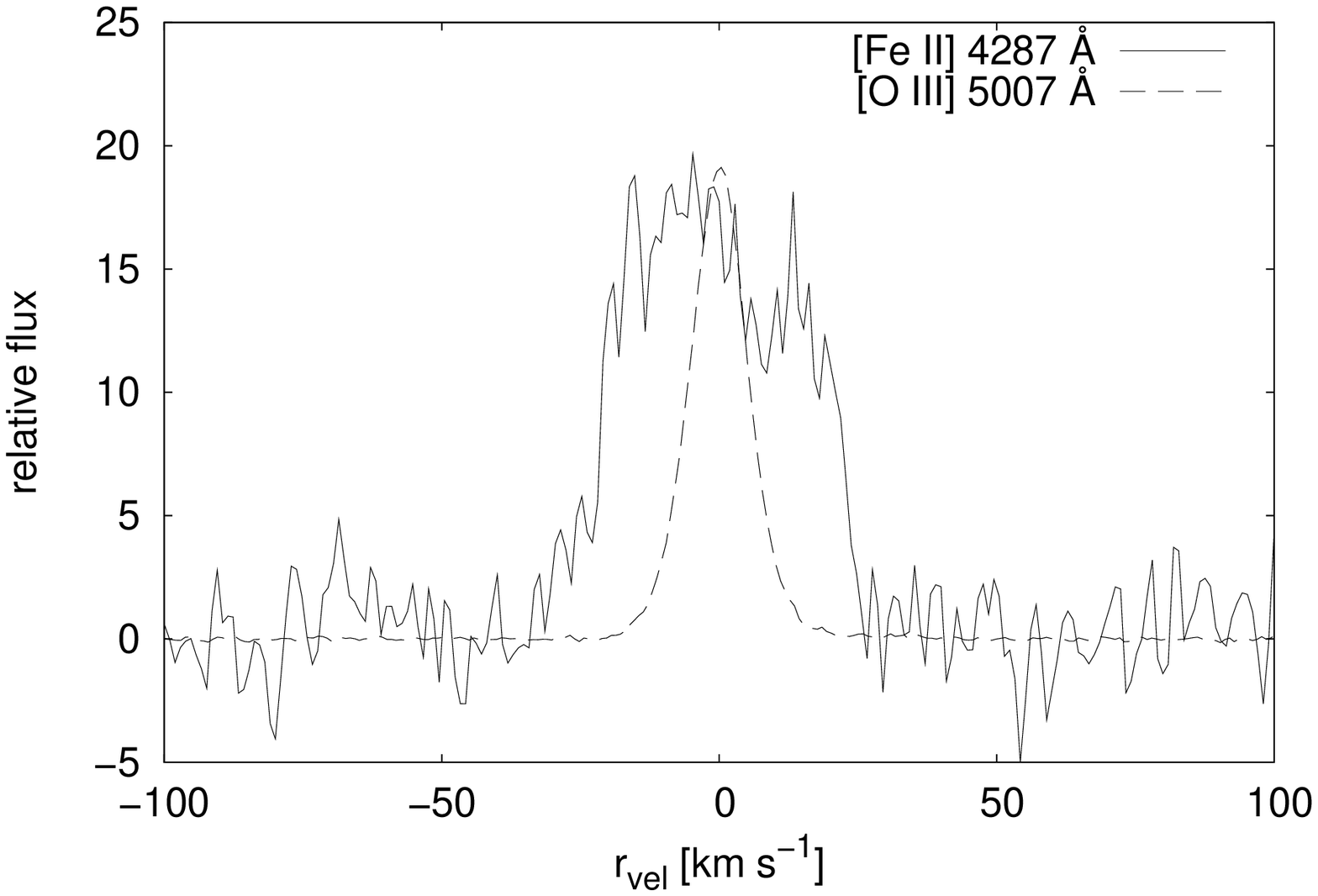}
\includegraphics[width=8cm]{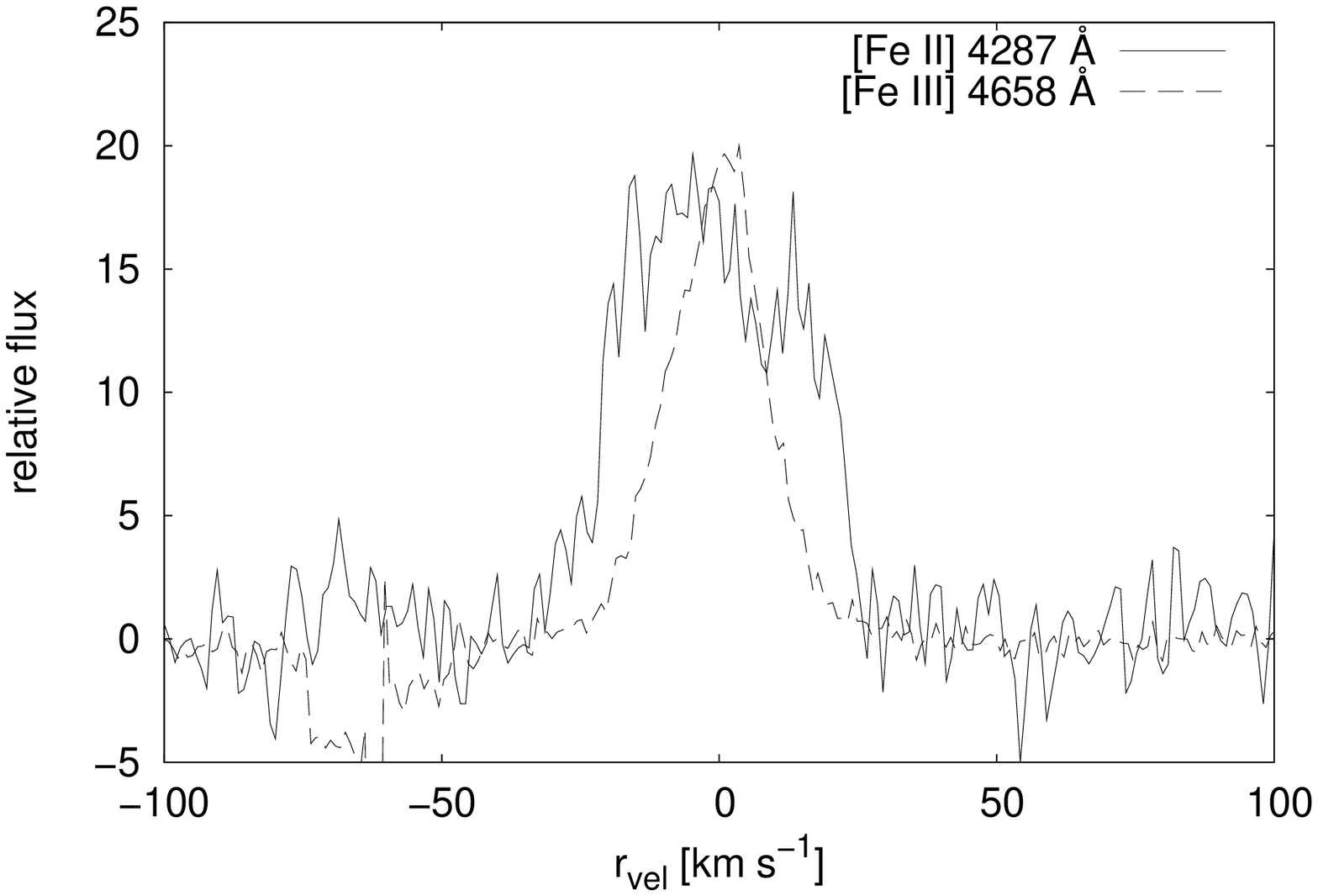}
   \caption{The VLT continuum subtracted spectrum of the [Fe\,{\sc ii}] 4287\,\AA\ line and the flux error (upper), [Fe\,{\sc ii}] 4287\,\AA\ and [O\,{\sc iii}] 5007\,\AA\ lines (middle) and [Fe\,{\sc ii}] 4287\,\AA\ and [Fe\,{\sc iii}] 4658\,\AA\ lines (lower).
           }
      \label{lines}
\end{figure}

The nebular spectrum is rich in [Fe\,{\sc ii}] and [Fe\,{\sc iii}] lines
(Table~4). The [Fe\,{\sc ii}], [O\,{\sc i}], [N\,{\sc i}] and permitted O\,{\sc
i} lines show wide profiles with sharp edges in the VLT spectrum. They can be
modelled with a shell expanding with a velocity of about $\rm 20\,km\,s^{-1}$.
[Fe\,{\sc iii}], [O\,{\sc ii}], and Si\,{\sc ii} lines show slighlty asymmetric
and narrower profiles (Figure \ref{lines}). Other nebular lines show symmetric
profiles.

\begin{figure}[]
\begin{center}
\includegraphics[width=8cm]{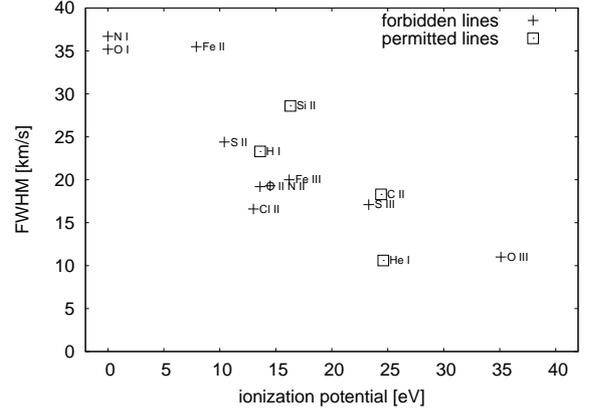}
\end{center}
\caption{Deconvolved line widths vs ionization potentials.}
\label{exp}
\end{figure}

The line widths for different ions against the ionization potential are
presented in the Figure \ref{exp}. The ions of higher ionization potential show
lower expansion velocity than the ions of lower ionization potential. As the
high excitation lines form closer to the central star than the low excitation
lines, the observed trend indicates the existence of a strong velocity gradient
across the nebular shell.

Stellar absorptions of H, He\,{\sc i}, and He\,{\sc ii} are present in the VLT
spectrum. Other stellar lines include O\,{\sc iii} and Si\,{\sc iv}
absorptions. No stellar wind lines were observed.

\subsection{Torun model}

\begin{figure*}
\begin{center}
\includegraphics[width=15cm]{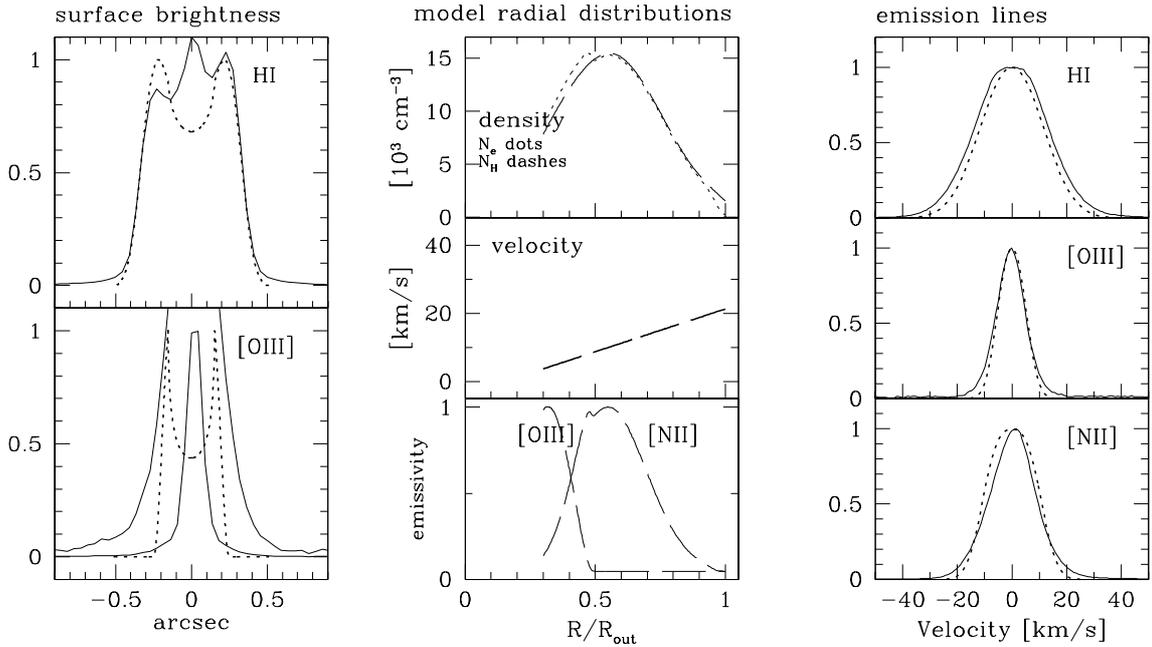}
\end{center}
\caption{The Torun model results for stellar temperature of
25,200\,K and luminosity of 1400\,$\rm L_\odot$. The observed data are
drawn in solid lines, while model fits and model parameters are shown
as dotted and dashed lines. The ``surface brightness'' panel shows the
slices through the H$\alpha$ and [\ion{O}{iii}] 5007\,\AA\ image centre taken
along a nebular short axis with the model brightness profile superposed.
Both slices are scaled to unity; however, the [\ion{O}{iii}] image is
additionally shown with different scaling (see the text) because the
centre is dominated by the stellar flux. The radial distributions of the best-fit model for selected parameters are shown next. The right-most panel shows the observed and modelled profiles of three nebular lines: H$\alpha$, [\ion{O}{iii}] 5007\,\AA, and [\ion{N}{ii}] 6584\,\AA.}
\label{best_fit_fig}
\end{figure*}

The object Hen\,2-260 is obviously a bipolar nebula; however, it is not
extremely elongated. We think of it as composed of a barrel-shaped main body
with moderate extensions (low density lobes) in the axial direction (which is
nearly in the plane of the sky). We propose to approximate this PN with a
spherical model, which certainly corresponds to the dominating dense barrel. We
expect that the lobes contribute a little to the total line fluxes, and by
considering the low spatial resolution of spectroscopy, we expect that skipping
the lobes is not a bad approximation.

We applied the Torun photo-ionization codes to de-compose the structure of the
nebula \citep{2003MNRAS.338..347G}. The star is assumed to be a black body. The
nebula is approximated as a spherical shell with a radial density distribution
that is similar to a reversed parabola and a velocity field that linearly
increases with radius. Locating the best-fit parameters is aided by a routine
with a genetic algorithm.

The code calculates 1-dimensional (radial) emissivity profiles. Assuming
spherical symmetry, they are converted to the observed projected intensity
profiles. These profiles are compared with the slices along the minor axis of
the HST images, which are averaged over 9 pixels. The model density distribution
is manually modified at the maximum, the inner radius and outer radius to obtain
a best similarity (judged visually) between both images.

With the derived density profile, a genetic algorithm searches for the stellar
parameters and for the nebular ionised mass that best fit the line flux ratios.
Later, the profiles of lines with different ionization states, tracing different
radii within the nebula are fitted to obtain the radial velocity profile within
the nebula.

It was relatively straightforward to find a density profile that reproduced the
H$\alpha$ slice down to the level of 10\% of the maximum intensity
(Fig.\,\ref{best_fit_fig}). The low density outer regions could be, in
principle, extended further out to reproduce the halo; however, the PN is more
complicated than a sphere, so we therefore did not go further with fine-tuning
of the density distribution. The density contrast between the maximum and the
halo is around 10. The [\ion{O}{iii}] line is very weak in Hen\,2-260, so we see
mainly the very strong stellar continuum in the HST image. The scaled-to-unity
image certainly is not appropriate to compare with the model. We show the
[\ion{O}{iii}] image scaled accordingly to its strengths relative to H$\alpha$
in Figure\,\ref{best_fit_fig}. The nebular emission is well localised in the
lowest intensity area.

The Torun model takes a limited number of most representative emission lines
into account. They are taken from SALT spectrum analysed in this paper. The
automated search routine found this fit for a stellar temperature of 25,200\,K
and a luminosity of 1400\,$\rm L_\odot$ with nebular ionised mass of 0.01\,$\rm
M_\odot$.

\subsection{Cloudy modelling}\label{cloudy}

\begin{table}
\caption{A list of observed/modelled parameters compared to the results of our Cloudy model of Hen\,2-260.}
\begin{center}
\begin{tabular}{ccc}
\hline
		&other authors	&this paper	\\
\hline
$\rm F_{WISE11.6\mu m} \, [mJy]$ & $640 \pm 9$	&	610	\\
$\rm F_{WISE22.1\mu m} \, [mJy]$ & $7540 \pm 50$	&	6820	\\
$\rm F_{IRAS12\mu m} \, [mJy]$ & $590 \pm 60$	&	649	\\
$\rm F_{IRAS25\mu m} \, [mJy]$ &$8100 \pm 500$	&	7370	\\
$\rm F_{IRAS60\mu m} \, [mJy]$ &$2900 \pm 300$	&	3322	\\
$\rm F_{IRAS100\mu m} \, [mJy]$ &$<4220$	&	\\
$\rm F_{AKARI65\mu m} \, [mJy]$ &$1399$		&	\\
$\rm F_{AKARI90\mu m} \, [mJy]$ &$1167 \pm 112$	&	1167	\\
$\rm F_{AKARI160\mu m} \, [mJy]$ &$210.3$	&	\\
$\rm F_{1.4\,GHz} \, [mJy]$	&$8.1 \pm 0.5$$^{\mathrm{a}}$&	\\
$\rm F_{5\,GHz} \, [mJy]$	&13$^{\mathrm{b}}$		&12.47	\\
${\rm log}\,F{\rm (H{\alpha}}) \, {\rm [erg\,cm^{-2}\,s^{-1}]}$&	$-11.26 \pm 0.04^{\mathrm{e}}$	&$-11.19 \pm 0.03$\\
${\rm log}\,F{\rm (H{\beta}}) \, {\rm [erg\,cm^{-2}\,s^{-1}]}$&	$-12.13 \pm 0.1^{\mathrm{h}}$	&	\\
\multirow{2}{*}{$\rm log (L/L_\odot)$}&$3.79^{\mathrm{c}}$&	3.11\\
{}&$4.09^{+0.15}_{-0.14} (3.8 \pm 0.26)^{\mathrm{d}}$&	\\
$\rm \in(N)$	&$7.14^{\mathrm{f}}, 7.24^{\mathrm{g}}$& 7.65	\\
$\rm \in(O)$	&$8.18^{\mathrm{f}}, 8.00^{\mathrm{g}}$& 8.69	\\
$\rm \in(S)$	&$6.29^{\mathrm{f}}, 6.22^{\mathrm{g}}$& 6.70	\\
$\rm \in(Ar)$	&$6.97^{\mathrm{f}}, 5.33^{\mathrm{g}}$& 5.99	\\
$\rm log(r_{in}) \, [cm]$&		&16.83	\\
$\rm log(r_{out}) \, [cm]$&		&17.20	\\
$\rm log {n}_H \, [cm^{-3}]$ &	&4.43	\\
$\rm T_{eff} [K]$ &$\rm 28,000^{\mathrm{i}}$	&35,000	\\

\hline
\label{parameters}
\end{tabular}
\end{center}
$^{\mathrm{a}}$\citet{1998ApJS..117..361C} $^{\mathrm{b}}$\citet{1990A&AS...84..229A} $^{\mathrm{c}}$for the adopted distance of 12\,kpc, \citet{2007A&A...467L..29G} $^{\mathrm{d}}$for the adopted distance of 8\,kpc, \citet{2007A&A...467.1253H} $^{\mathrm{e}}$\citet{2013MNRAS.431....2F} $^{\mathrm{f}}$\citet{2004A&A...414..211E} $^{\mathrm{g}}$\citet{2009A&A...494..591C} $^{\mathrm{h}}$\citet{1992secg.book.....A}
$^{\mathrm{i}}$\citet{2007A&A...467.1253H}
\end{table}

\begin{table}
\caption{Observed, dereddened, and modelled line fluxes of Hen\,2-260 obtained with the SALT and {\it 1.9\,m Radcliffe} telescopes that are normalised relative to $\rm F(H\beta)=100$.}
\begin{center}
\begin{tabular}{ccccc}
\hline
$\lambda_{lab}$ [\AA]	&ion	&observed&dereddened	&modelled\\
\hline
3726.03+3728.82	&[O\,{\sc ii}]	&{\it 69.1}&{\it 109.7}	&120.5	\\
3835.38		&H\,{\sc i}	&{\it 3.7}&{\it 5.7}  	&7.8	\\
3889.05		&H\,{\sc i}	&{\it 6.9}&{\it 10.2} 	&10.6	\\
3970.07		&H\,{\sc i}	&{\it 10.2}&{\it 14.6} 	&16.3	\\
4068.60		&[S\,{\sc ii}]	&{\it 4.5}&{\it 6.2}  	&5.3	\\
4076.01		&[S\,{\sc ii}]	&{\it 2.4}&{\it 3.3}  	&1.8	\\
4101.74		&H\,{\sc i}	&{\it 17.0}&{\it 23.1} 	&27.2	\\
4340.46		&H\,{\sc i}	&{\it 36.4}&{\it 45.3} 	&48.6	\\
4658.05		&[Fe\,{\sc iii}]&2.00	&2.19  	&0.51	\\
4701.54         &[Fe\,{\sc iii}]&0.88	&0.94  	&0.22	\\
4733.90         &[Fe\,{\sc iii}]&0.37	&0.39  	&0.10	\\
4754.69         &[Fe\,{\sc iii}]&0.34	&0.35  	&0.09	\\
4769.43         &[Fe\,{\sc iii}]&0.24	&0.25  	&0.07	\\
4777.68         &[Fe\,{\sc iii}]&0.18	&0.19  	&0.05	\\
4814.53         &[Fe\,{\sc ii}]	&0.18	&0.18	&	\\
4861.33         &H\,{\sc i}	&100.00	&100.00	&100.00	\\
4881.00		&[Fe\,{\sc iii}]&1.15	&1.14  	&0.27	\\
4930.54		&[Fe\,{\sc iii}]&0.13	&0.13  	&0.03	\\
4958.91         &[O\,{\sc iii}]	&2.41	&2.33  	&2.23	\\
4987.21		&[Fe\,{\sc iii}]&0.12	&0.11  	&0.05	\\
5006.84         &[O\,{\sc iii}]	&7.85	&7.45  	&6.71	\\
5015.68		&He\,{\sc i}	&0.50	&0.47  	&0.45	\\
5041.02         &Si\,{\sc ii}	&0.28	&0.26  	&	\\
5055.98         &Si\,{\sc ii}	&0.48	&0.44  	&	\\
5084.77		&[Fe\,{\sc iii}]&0.07	&0.06  	&0.02	\\
5146.58         &O\,{\sc i}	&0.19	&0.17  	&	\\
5158.78         &[Fe\,{\sc ii}]	&0.38	&0.33  	&	\\
5197.90         &[N\,{\sc i}]	&0.36	&0.31  	&0.21	\\
5261.62         &[Fe\,{\sc ii}]	&0.22	&0.18  	&	\\
5270.40         &[Fe\,{\sc iii}]&1.54	&1.30  	&0.31	\\
5299.01         &O\,{\sc i}	&0.25	&0.21  	&	\\
5333.65         &[Fe\,{\sc ii}]	&0.10	&0.08  	&	\\
5512.73         &O\,{\sc i}	&0.18	&0.14  	&	\\
5517.72         &[Cl\,{\sc iii}]&0.07	&0.06  	&0.05	\\
5527.61         &[Fe\,{\sc ii}] &0.07	&0.05  	&	\\
5537.89         &[Cl\,{\sc iii}]&0.20	&0.15  	&0.13	\\
5554.83         &O\,{\sc i}	&0.22	&0.16  	&	\\
5695.19		&C\,{\sc iii}	&0.21	&0.15  	&	\\
5754.65         &[N\,{\sc ii}]	&4.19	&3.02  	&2.91	\\
5875.66          &He\,{\sc i}	&2.89	&2.03  	&2.31	\\
5958.51         &O\,{\sc i}	&0.40	&0.27  	&	\\
5978.93          &Si\,{\sc ii}	&0.24	&0.16  	&	\\
6046.39         &O\,{\sc i}	&0.54	&0.36  	&	\\
6300.30         &[O\,{\sc i}]	&3.38	&2.10  	&2.14	\\
6312.10         &[S\,{\sc iii}]	&1.49	&0.92  	&1.28	\\
6347.09         &Si\,{\sc ii}	&0.57	&0.35  	&	\\
6363.78         &[O\,{\sc i}]	&1.12	&0.68  	&0.68	\\
6371.38		&Si\,{\sc ii}	&0.32	&0.19  	&	\\
6548.10         &[N\,{\sc ii}]	&62.87	&36.23 	&38.86	\\
6562.77         &H\,{\sc i}	&493.25	&283.24	&282.50	\\
6583.50         &[N\,{\sc ii}]	&185.90	&106.22	&114.68	\\
6666.80         &[Ni\,{\sc ii}]	&0.08	&0.05  	&	\\
6678.15         &He\,{\sc i}	&0.80	&0.45  	&0.63	\\
6716.44         &[S\,{\sc ii}]	&3.58	&2.00  	&1.66	\\
6730.82         &[S\,{\sc ii}]	&8.00	&4.46  	&3.67	\\
7002.21         &O\,{\sc i}	&0.72	&0.38  	&	\\
7065.25         &He\,{\sc i}	&1.92	&1.00  	&0.77	\\
7135.8          &[Ar\,{\sc iii}]&2.29	&1.17  	&1.22	\\
7155.16         &[Fe\,{\sc ii}]	&0.73	&0.37  	&	\\
7254.53         &O\,{\sc i}	&0.81	&0.41  	&	\\
7319.46         &[O\,{\sc ii}]	&49.78	&24.27 	&21.86	\\
7330.13         &[O\,{\sc ii}]	&40.44	&19.67 	&17.72	\\
7377.83         &[Ni\,{\sc ii}]	&0.90	&0.44  	&	\\
7388.18		&[Fe\,{\sc ii}]	&0.19	&0.09  	&	\\
7411.61		&[Ni\,{\sc ii}]	&0.16	&0.08  	&	\\
\hline
\label{ratiossalt}
\end{tabular}
\end{center}
\end{table}

We computed a grid of models to fit the observed fluxes using Cloudy c10.01, as
described by \citet{2013RMxAA..49..137F}. The distance of 12\,kpc was adopted.
We used a spherical model of the nebula with a constant density. Central star
radiation was modelled with TLUSTY model atmospheres
\citep{1995ApJ...439..875H}. The stellar luminosity, temperature, inner radius,
nebular density, and elemental and grain abundances were varied to find the best
solution \citep{1999MNRAS.308..623V}. The model used silicate grains. We used
the $\rm 90 \, \mu m$ flux as a stopping criterion.

Initially, we did not include the full list of the singly and doubly ionised
iron transitions. We added the iron lines to the model after the solution
converged. All the observational constraints used are given in the Table
\ref{parameters}. The full list of line fluxes obtained from the SALT spectrum
and modelled with the Cloudy is given in the Table \ref{ratiossalt}.

There is a severe discrepancy in the elemental abundances obtained by other
authors and us (Table \ref{parameters}). \citet{2009A&A...494..591C} report
large uncertainties for the determination of nebular abundances for Hen\,2-260
due to poorly known ionizing correction factors for this object. The V-band
magnitude of 13.4 (assuming the distance of 12\,kpc) is a bit fainter than the
observed brightness (Figure \ref{hst}), taking the extinction into account. 
Some differences between the observed and modelled fluxes for individual lines
exceed the observational uncertainties, but all the diagnostic line ratios agree
very well with the ratios predicted by the model.

After the Cloudy models converged to the best solution, we ran a grid of models
for different temperatures, which ranged from 34.6 to 35.8\,kK to fit the
temporal evolution of the [O\,{\sc iii}] 5007\,\AA\ line flux. We have chosen
this line because it is not affected by stellar absorptions, it is stronger than
other high excitation lines, and it is very close to the $\rm H \beta$
line, so the flux ratio of both lines is not very sensitive to the
wavelength-dependent calibration errors, such as atmospheric extinction
correction or differential atmospheric refraction. We also scaled the nebular
density and inner radius, assuming a constant expansion rate.

The observed [O\,{\sc iii}] 5007\,\AA\ flux was best modelled with the stellar
temperature of $\rm 34875 \pm 91\,K$ in 2001, $\rm 34944 \pm 20\,K$ in 2005, and
$\rm 35289 \pm 52\,K$ in 2012. The temperature uncertainties correspond only to
the standard deviation of the [O\,{\sc iii}] 5007\,\AA\ line flux. Systematic
error is larger, but it is similar in all the temperature determinations and is
cancelled out when computing the difference between the temperatures. Thus, the
heating rate can be firmly established.

\begin{figure}[]
\begin{center}
\includegraphics[width=8cm]{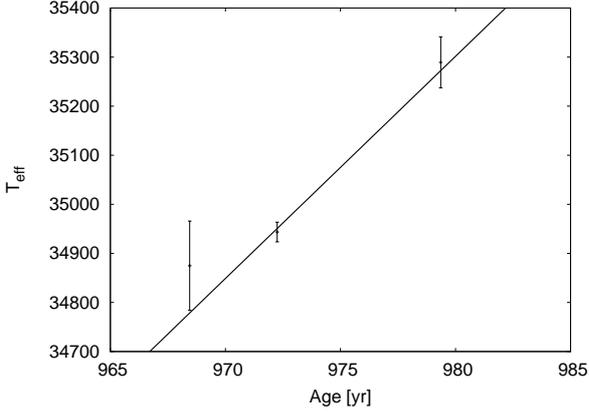}
\end{center}
\caption{The observed temperature evolution of the central star. The solid line represents the evolutionary track for the interpolated model with final mass of $\rm 0.626M_{\odot}$ \citep{1995A&A...299..755B}.}
\label{tempevol}
\end{figure}

The observed heating rate of $\rm 45 \pm 7\,K\, \rm yr^{-1}$ can be modelled
with the post-AGB evolutionary track that is interpolated for the final mass of
$\rm 0.626 ^{+0.003} _{-0.005} \, M_{\odot} \, (0.626 ^{+0.007} _{-0.010} \,
M_{\odot})$ by using the models published by \citet{1995A&A...299..755B} or $\rm
0.645 ^{+0.008} _{-0.008} \, M_{\odot} \, (0.645 ^{+0.015} _{-0.016} \,
M_{\odot})$ by using  \citet{1993ApJ...413..641V} at a one (two) sigma
confidence level (Figure \ref{tempevol}).

The mass errors correspond to the heating rate uncertainty. The mass of the star
is very close to the peak of the mass distribution for central stars of
planetary nebulae of $\rm 0.61 \, M_{\odot}$ \citep{2007A&A...467L..29G}. The
age of the central, spherical component of the nebula of the order of 1000 years
(for a distance of 12\,kpc, $\rm r_{out} = 0.25 \,arcsec$ and $\rm V_{\exp} =
20\,km\,s^{-1}$) is in better agreement with the post-AGB evolutionary track
value given by Bl\"{o}cker (980 years) than Vassiliadis \& Wood (480\,yr).
However, the age of the object determined from the evolutionary tracks depends
on the limited accuracy of the absolute temperature determination of the star.
If we adopted 28,000\,K instead of about 35,000\,K, which is modelled with
Cloudy, then the derived age would be reduced by about 200\,yr. On the other
hand, the age depends on the adopted zero scale by the evolutionary models.

\begin{figure}[]
\begin{center}
\includegraphics[width=8cm]{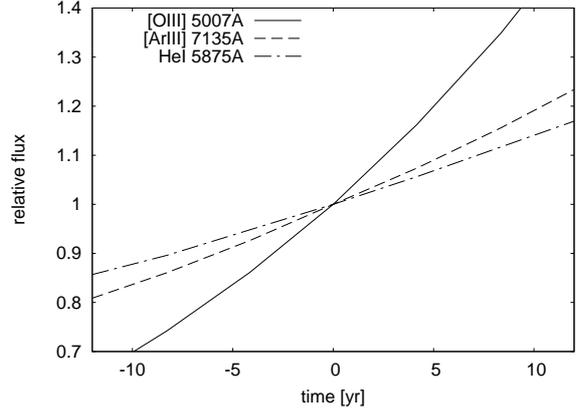}
\end{center}
\caption{The predicted evolution of three line fluxes. The zero point on the time axis corresponds to 2012 (SALT observation of the He\,2-260).}
\label{fluxevol}
\end{figure}

The Cloudy models predict a 16\% flux increase for the He\,{\sc i} 5875\,\AA\
line, a 13\% increase for the He\,{\sc i} 6678\,\AA\ line, a 20\% increase for
the He\,{\sc i} 7065\,\AA\ line, and a 18\% increase for the [Ar\,{\sc iii}]
7135\,\AA\ line between 2001 and 2012 (Figure \ref{fluxevol}). We observe a
faster flux increase for those lines. However, those lines are fainter than the
[O\,{\sc iii}] 5007\,\AA\ line, and some of them can be affected by the stellar
absorptions, so the flux change determination is less reliable.

\begin{figure}[]
\begin{center}
\includegraphics[width=8cm]{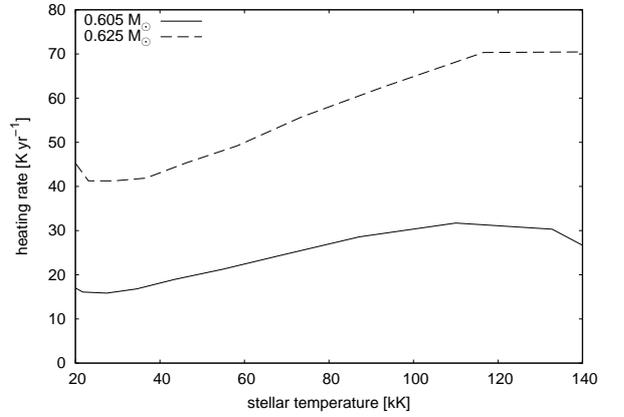}
\end{center}
\caption{The modelled evolution of the heating rate of central stars for two different masses \citep{1995A&A...299..755B}.}
\label{heating}
\end{figure}

A significant systematic error may affect the determination of the absolute
central star temperature. However, the heating rate depends primarily on the
central star mass and does not change steeply during the evolution of a star
between 20,000 and 140,000\,K (Figure \ref{heating}). The mass derivation based
on the heating rate does not depend on the absolute age of the nebula, unlike
the dynamical method used by \citet{2007A&A...467L..29G}.

\subsection{Optical photometry}

\begin{figure*}[]
\begin{center}
\includegraphics[width=14cm]{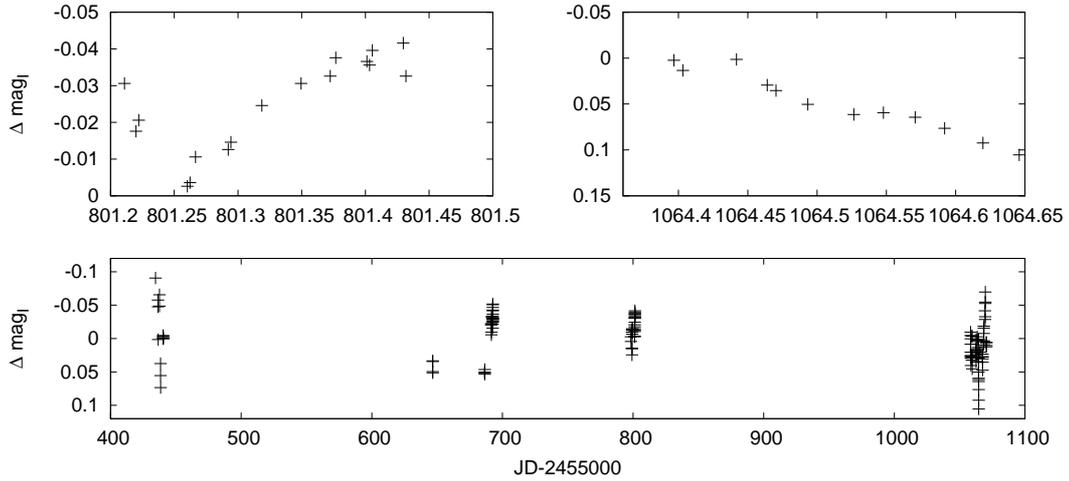}
\end{center}
\caption{SAAO I band 2010-2012 photometry of Hen\,2-260. Photometry taken at two separate nights is shown above the lighcurve collected between 2010 and 2012.}
\label{photometry}
\end{figure*}

The central star of the PN Hen\,2-260 shows photometric variability on a
timescale of hours or days (Figure \ref{photometry}). Photometry taken at
individual nights cannot be clearly seen in the bottom of Figure
\ref{photometry} due to large time range of the observations shown. Two best
sampled individual nights are plotted at the top of Figure \ref{photometry}. The
observations do not indicate sinusoidal variations. The amplitude of the
variability is of the order of 0.1 mag in the I band. However, the real
amplitude may be somewhat higher, since the nebular contribution was not
removed.

\begin{figure}[]
\begin{center}
\includegraphics[width=8cm]{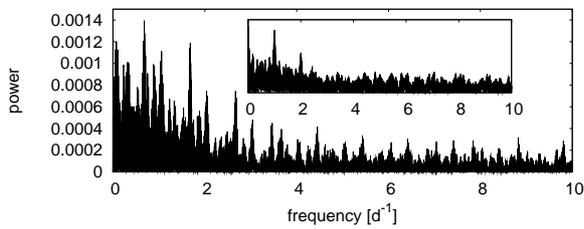}
\end{center}
\caption{Power spectrum of I-band observations of Hen\,2-260 along with the power window shown in the inset.}
\label{periodogram}
\end{figure}

No single dominant periodicity can be identified in the power spectrum (Figure
\ref{periodogram}). The power spectrum is noisy; however, two strongest peaks
can be identified at 0.68 and $\rm 1.68\,d^{-1}$. The latter one may be the
alias of the former. Another peak is present at $\rm 1.06\,d^{-1}$.

The photometric data were taken non-uniformly during different seeing
conditions. We cannot draw firm conclusions regarding the nature of the
photometric variability of Hen\,2-260. It could be caused by pulsations,
magnetic activity of the star, binarity, or a combination of those.

We do not detect any emission lines originating from the stellar wind in the VLT
spectrum; thus, the variability of the central star can be most likely
attributed to stellar pulsations. Our object fulfills the criteria of a group of
variable CSPNs \citep{2003IAUS..209..237H}: the temperature between 25 and
50\,kK, (semi)regular variability on a timescale of several hours and
hydrogen-rich spectra. However, more uniform coverage that is compatible with
the variability timescale of the object would be needed to constrain the nature
of the variability.

\section{Discussion}

The spectra of most of PNe are usually observed only once in a lifetime. Most of
the PNe included in the largest catalogue of the PN fluxes to date, which are
provided by \citet{1992secg.book.....A}, have never been observed again. Our
results show that monitoring of nebular fluxes may be used as an excellent tool
to study the evolution of central stars of PNe. Spectroscopic observations of a
PN that cover a time span of only a decade can already reveal the temperarure
change of its central star. The derived heating rate of central stars is steeply
dependent on the mass of the star. The comparison of the heating rate with the
post-AGB evolutionary models can provide precise mass determinations. 

We discovered a 50\% increase of the [O\,{\sc iii}] 5007\,\AA\ emission line
flux in a period of eleven years in Hen\,2-260. The pace of the evolution of the
central star and the kinematical age of the PN are consistent with the existing
evolutionary models. Further observations will allow us to determine the heating
rate and mass of the star more accurately.

\begin{acknowledgements}

This work was financially supported by NCN of Poland through grants No.
2011/01/D/ST9/05966, 2011/03/B/ST9/02552, N\,N203\,511838 and 719/N-SALT/2010/0.
PvH acknowledges support from the Belgian Science Policy office through the ESA
PRODEX program. This paper uses observations made at the South African
Astronomical Observatory (SAAO). Some of the observations reported in this paper
were obtained with the Southern African Large Telescope (SALT), proposal
2011-3-POL-002, P.I. M. Hajduk. We made use of the Atomic Line List available at http://www.pa.uky.edu/$\sim$peter/newpage/.

\end{acknowledgements}

\bibliographystyle{aa} 
\bibliography{he2260}

\end{document}